\documentclass[a4paper]{jpconf}

\usepackage{graphicx}
\usepackage{graphicx}
\usepackage{graphicx}
\usepackage{bm}
\usepackage{amssymb}
\usepackage{amsmath}
\def\be{\begin{equation}}
\def\ee{\end{equation}}
\def\bed{\begin{displaymath}}
\def\eed{\end{displaymath}}
\def\bea{\begin{eqnarray}}
\def\eea{\end{eqnarray}}
\def\bear{\begin{array}}
\def\eear{\end{array}}
\def\bes{\begin{subequations}}
\def\ees{\end{subequations}}
\newcommand{\A}{{\mathcal{A}}}
\newcommand{\tA}{{\widetilde {\mathcal{A}}}}

\newcommand{\MSbar}{\overline{\rm MS}}  

\newcommand{\bL}{{\overline {\Lambda}}}

\newcommand{\tk}{{\widetilde k}}

\usepackage{graphics}
\usepackage{graphicx}
\usepackage{dcolumn}
\usepackage{bm}
\usepackage{epsfig}
\usepackage{graphicx}
\usepackage{multirow}
\usepackage{dcolumn}
\usepackage{graphicx,epsfig}%

\def\lsim{\mathrel{\rlap{\lower4pt\hbox{\hskip1pt$\sim$}}
    \raise1pt\hbox{$<$}}}         
\def\gsim{\mathrel{\rlap{\lower4pt\hbox{\hskip1pt$\sim$}}
    \raise1pt\hbox{$>$}}}         

\begin{document}
\title{Mathematica and Fortran programs for various analytic QCD couplings}\footnote{Preprint USM-TH-330. Based on the presentation given by G.C. at the 16th International workshop on Advanced Computing and Analysis Techniques in physics research (ACAT 2014), Prague, Czech Republic, September 1-5, 2014. To appear in the proceedings by the IOP Conference Series publishing.}
\author{C\'esar Ayala and 
Gorazd Cveti\v{c}}
\address{Department of Physics, Universidad T\'ecnica Federico Santa Mar\'{\i}a, Casilla 110-V, Valpara\'iso, Chile}
\ead{gorazd.cvetic@gmail.com}

\begin{abstract}
We outline here the motivation for the existence of analytic QCD models, i.e., QCD frameworks in which the running coupling $\A(Q^2)$ has no Landau singularities. The analytic (holomorphic) coupling $\A(Q^2)$ is the analog of the underlying pQCD coupling $a(Q^2) \equiv \alpha_s(Q^2)/\pi$, and any such $\A(Q^2)$ defines an analytic QCD model. We present the general construction procedure for the couplings $\A_{\nu}(Q^2)$ which are analytic analogs of the powers $a(Q^2)^{\nu}$. Three analytic QCD models are presented. Applications of our program (in Mathematica) for calculation of $\A_{\nu}(Q^2)$ in such models are presented. Programs in both Mathematica and Fortran can be downloaded from the web page: gcvetic.usm.cl.
\end{abstract}

\section{Why analytic QCD?}
\label{sec:why}

Perturbative QCD (pQCD) running coupling $a(Q^2)$ 
[$\equiv  \alpha_s(Q^2)/\pi$, where
$Q^2 \equiv - q^2$]
has  unphysical (Landau) singularities
at low spacelike momenta $0 < Q^2 \stackrel{<}{\sim} 1 \ {\rm GeV}^2$.

For example, the one-loop pQCD running coupling
{\small
\be
a(Q^2)^{(1-\ell.)} = \frac{1}{\beta_0 \ln (Q^2/\Lambda^2_{\rm Lan.})}
\ee
}
has a Landau singularity (pole) at 
$Q^2 = \Lambda^2_{\rm Lan.}$ ($ \sim 0.1 \ {\rm GeV}^2$). 
The 2-loop pQCD coupling $a(Q^2)^{(2-\ell.)}$ 
has a Landau pole at $Q^2 = \Lambda^2_{\rm Lan.}$ and Landau cut at $0< Q^2 < \Lambda^2_{\rm Lan.}$.

It is expected that the true QCD coupling ${\A}(Q^2)$ has no such singularities. Why? 

General principles of QFT dictate that any spacelike observable ${\cal D}(Q^2)$ (correlators of currents, structure functions, etc.) is an analytic (holomorphic) function of $Q^2$ in the entire $Q^2$  complex plane with the exception of the timelike axis: 
$Q^2 \in \mathbb{C} \backslash (-\infty, -{M^2_{\rm thr.}}]$,
where ${M_{\rm thr.}} \sim 0.1$ GeV is a threshold  mass ($\sim M_{\pi}$).
If ${\cal D}(Q^2)$ can be evaluated as a leading-twist term, then it is a function of the running coupling $a(\kappa Q^2)$ where $\kappa \sim 1$:
${\cal D}(Q^2) = {\cal F}(a(\kappa Q^2))$.
Then the argument $a(\kappa Q^2)$ is expected to have the same
analyticity properties as ${\cal D}$, which is { not}
the case with the pQCD coupling in the usual renormalization schemes ($\MSbar$,
't Hooft, etc.).

A QCD coupling ${\A}(Q^2)$ with {holomorphic} behavior for
$Q^2 \in \mathbb{C} \backslash (-\infty, -{M^2_{\rm thr.}}]$, represents an  {analytic QCD} model ({anQCD}).

Such holomorphic behavior comes usually together
with ({IR-fixed-point}) behavior [${\A}(0) < \infty$].
The {IR-fixed-point} behavior of ${\A}(Q^2)$ is suggested by:
\begin{itemize}
\item
{lattice} calculations \cite{latt1,latt2,latt3};
calculations based on Dyson-Schwinger equations (DSE) \cite{DSE1,DSE2};
Gribov-Zwanziger approach \cite{GZ1,GZ2}; 
\item
The {holomorphic} ${\A}(Q^2)$ with {IR-fixed-point} behavior was proposed in various {analytic QCD} models, among them: 
\begin{enumerate}
\item
Analytic Perturbation Theory (APT) of Shirkov, Solovtsov et al. \cite{ShS1,ShS2,Mil,Sh,KS};
\item
its extension {Fractional APT (FAPT)} \cite{BMS1,BMS2,BMS3};
\item
{analytic models} with ${\A}(Q^2)$ very close to $a(Q^2)$ at high $|Q^2| > {\Lambda^2_{\rm Lan.}}$:
${\A}(Q^2) - a(Q^2) \sim ({\Lambda^2_{\rm Lan.}}/Q^2)^{N}$ with ${N=3, 4}$ or ${5}$, \cite{Webber,Alekseev,CCME,2danQCD};
\item
{Massive Perturbation Theory (MPT)}, \cite{Simonov1,Simonov2,BadKuz,ShMPT}.
\end{enumerate}
\end{itemize}

{Perturbative QCD (pQCD)} 
can give {analytic} coupling $a(Q^2)$ in specific schemes
with IR fixed point; the condition of reproduction of the correct value
of the (strangeless and massless) semihadronic $\tau$ lepton $V+A$ decay ratio 
${r_{\tau} \approx 0.20}$ strongly restricts such schemes
\cite{anpQCD1,anpQCD2,anpQCD3}.


If the {analytic} coupling ${\A}(Q^2)$ is 
{not perturbative},
${\A}(Q^2)$ {differs} from the pQCD
couplings $a(Q^2)$ at $|Q| \stackrel{>}{\sim} 1$ GeV 
by { nonperturbative (NP) terms}, typically by
some power-suppressed terms $\sim 1/Q^{2 N}$ or 
$1/[Q^{2 N} \ln^K(Q^2/{\Lambda^2_{\rm Lan.}})]$.

An analytic QCD model which gives $\A(0) = \infty$ was constructed in
\cite{Nest1a,Nest1b,Nest1c}.

\section{The formalism of constructing $\A_{\nu}$ in general anQCD}
\label{sec:formAnu}

Having ${\A}(Q^2)$ 
[the analytic analog of $a(Q^2)$] specified,
we want to evaluate the physical QCD quantities ${\cal D}(Q^2)$
in terms of such ${\A}(\kappa Q^2)$.

Usually ${\cal D}(Q^2)$ is known as a
(truncated) {power} series in terms of the pQCD
coupling $a(\kappa Q^2)$:
{\small
\be
{\cal D}(Q^2)^{[{N}]}_{\rm pt} =  a(\kappa Q^2)^{{\nu_0}} + 
d_1(\kappa) a(\kappa Q^2)^{{\nu_0}+1} + \ldots 
+ d_{N-1}(\kappa) a(\kappa Q^2)^{{\nu_0}+{N}-1}.
\label{pt}
\ee
}
In anQCD, the simple replacement $a(\kappa Q^2)^{{\nu_0}+m} \mapsto 
{\A}(Q^2)^{{\nu_0}+m}$ is {not} correct, it leads to a strongly diverging
series when ${N}$ increases, as argued in \cite{GCtech}; a different
formalism was needed, and was developed for general anQCD, first for 
the case of integer ${\nu_0}$ \cite{CV1,CV2},
and then for the case of general ${\nu_0}$ \cite{GCAK}.
It results in the replacements
{\small
\be
a(\kappa Q^2)^{{\nu_0}+m} \mapsto 
{\A}_{{\nu_0}+m}(Q^2)
\quad \left[ \not= {\A}(Q^2)^{{\nu_0}+m} \right] \ ,
\label{an1}
\ee
}
where the construction of the analytic power analogs
${\A}_{{\nu_0}+m}(Q^2)$
from ${\A}(Q^2)$ was obtained.

The construction starts with logarithmic derivatives of ${\A}(Q^2)$
[where $\beta_0=(11-2 N_f/3)/4$]:
{\small
\be
{\tilde{\mathcal{A}}}_{n+1}(Q^2) \equiv \frac{(-1)^n}{\beta_0^n n!}
\left( \frac{ \partial}{\partial \ln Q^2} \right)^n
{\A}(Q^2) \ , \qquad (n=0,1,2,\ldots) \ ,
\label{tAn}
\ee
}
and ${\tA}_1 \equiv {\A}$. Using the Cauchy theorem, these quantities can
be expressed in terms of the {discontinuity} function of anQCD coupling
${\tA}$ along its cut, ${\rho}(\sigma) \equiv {\rm Im} {\A}(-\sigma - i \epsilon)$
{\small
\be
{\tA}_{n+1}(Q^2) = \frac{1}{\pi} \frac{(-1)}{\beta_0^n \Gamma(n+1)}
\int_{0}^{\infty} \ \frac{d \sigma}{\sigma} {\rho}(\sigma)  
{\rm {Li}}_{-n} ( -\sigma/Q^2 ) \ .
\label{disptAn2}
\ee
}
This construction can be extended to a general
noninteger $n \mapsto {\nu}$
{\small
\be
{\tA}_{{\nu}+1}(Q^2) = \frac{1}{\pi} \frac{(-1)}{\beta_0^{{\nu}} \Gamma({\nu}+1)}
\int_{0}^{\infty} \ \frac{d \sigma}{\sigma} {\rho}(\sigma)  
{\rm {Li}}_{-{\nu}}\left( - \frac{\sigma}{Q^2} \right) \quad (-1 < {\nu}) \ .
\label{tAnu1}
\ee
}
This can be recast into an alternative form, involving ${\A}$ ($\equiv \tA_1$) instead of
${\rho}$
{\small
\bea
{\tilde{\mathcal{A}}}_{{\delta}+m}(Q^2) & = &
K_{{\delta}, m} 
\left(\frac{d}{d \ln Q^2}\right)^{m}
\int_0^1 \frac{d \xi}{\xi} {\mathcal{A}}(Q^2/\xi) \ln^{-{\delta}}\left(\frac{1}{\xi}\right) \ ,
\label{tAnu2}
\eea
}
where: $0\leq {{\delta}} <1$ and $m=0,1,2,\ldots$; $K_{{\delta},m} =(-1)^m \beta_0^{-{\delta}-m+1}/[\Gamma({\delta}+m)\Gamma(1-{\delta})]$.
This expression was obtained from Eq.~(\ref{tAnu1}) by the use of the
following expression for the ${\rm Li}_{-\nu}(z)$ function 
\cite{KKSh}:
{\small
\be
{\rm {Li}}_{-{n}-{\delta}}(z) = \left( \frac{d}{d \ln z} \right)^{{n}+1}
\left[ \frac{z}{\Gamma(1 - {\delta}) } \int_0^1 \frac{d \xi}{1 - z \xi}
\ln^{-{\delta}} \left( \frac{1}{\xi} \right) \right] 
\quad ({n} =-1, 0, 1, \ldots; 0 < {{\delta}} < 1) \ .
\label{Li_nugen}
\ee
}

The analytic analogs ${\A}_{{\nu}}$ of powers $a^{{\nu}}$ are then obtained by
combining various generalized logarithmic derivatives
(with the coefficients ${\tk}_m({\nu})$ obtained in \cite{GCAK})
{\small
\be
{\A}_{\nu}={\tA}_{{\nu}}
+\sum_{m\geq1} {\tk}_m({\nu}) {\tA}_{{\nu}+m} \ .
\label{AnutAnu}
\ee
}

\section{The considered anQCD models}
\label{sec:anQCDmod}

We constructed Mathematica and Fortran programs for three anQCD models:
1.) {Fractional Analytic Perturbation Theory (FAPT)} \cite{BMS1,BMS2,BMS3};
2.) {2$\delta$ analytic QCD (2$\delta$anQCD)} \cite{2danQCD};
3.) {Massive Perturbation Theory (MPT)} \cite{Simonov1,Simonov2,BadKuz,ShMPT}.
These three models are described below.

\subsection{anQCD models: FAPT}
\label{subs:FAPT}

\noindent
Application of the Cauchy theorem to the function $a(Q^{'2})^{{\nu}}/(Q^{'2}-Q^2)$
gives
{\small
\begin{equation}
a(Q^2)^{{\nu}} = \frac{1}{\pi} \int_{\sigma= - {{\Lambda^2_{\rm Lan.}}} - \eta}^{\infty}
\frac{d \sigma {\rm Im}(a(-\sigma - i \epsilon)^{{\nu}})}{(\sigma + Q^2)},
   \quad (\eta \to +0).
\label{adisp}
\end{equation}
}
In {FAPT}, the integration over the Landau part of the cut in the above integral
is eliminated; since  $\sigma \equiv -Q^2$, the Landau cut is $-{{\Lambda^2_{\rm Lan.}}} < \sigma < 0$.
This leads to the {FAPT} coupling
{\small
\begin{equation}
{{\A}}^{\rm {{(FAPT)}}}_{{\nu}}(Q^2) = \frac{1}{\pi} \int_{\sigma= 0}^{\infty}
\frac{d \sigma  {\rm Im}(a(-\sigma - i \epsilon)^{{\nu}})
}{(\sigma + Q^2)} \ .
\label{AnuFAPT}
\end{equation}
}

\subsection{anQCD models: 2$\delta$QCD}
\label{subs:2danQCD}

\noindent
Here, $ {{\rho}}(\sigma) \equiv {\rm Im}  {\A}(-\sigma - i \epsilon)$
is approximated at {high} momenta 
$\sigma \geq {M_0^2}$ by  ${\rho^{\rm (pt)}}(\sigma)$ [$\equiv {\rm Im} \  a(-\sigma - i \epsilon)$], and in the unknown {low}-momentum regime by {two deltas}:
{\small
\be
{\rho}^{({2 \delta})}(\sigma) =  \pi {F_1^2} \delta(\sigma - {M_1^2})
+ \pi {F_2^2} \delta(\sigma - {M_2^2}) + \Theta(\sigma-{M_0^2}) {\rho}^{\rm (pt)}(\sigma) \quad \Rightarrow
\label{rho2d}
\ee
\be
{\tA}_{{\nu}}^{({2 \delta})}(Q^2) = 
\frac{(-1)}{\beta_0^{\nu} \Gamma(\nu\!+\!1)} {\bigg\{}
\sum_{{j=1}}^{{2}} 
\frac{{F_j^2}}{{M_j^2}}
{\rm {Li}}_{-\nu}\left( - \frac{{M_j^2}}{Q^2} \right)  
+ \frac{1}{\pi} 
\int_{{M_0^2}}^{\infty} \ \frac{d \sigma}{\sigma} {\rm Im} a 
(-\sigma\!-\!i \epsilon) 
{\rm {Li}}_{-\nu}\left( - \frac{\sigma}{Q^2} \right) {\bigg\}}.
\label{tAnu2d} 
\ee
}
The parameters ${F_j^2}$ and ${M_j}$ ($j=1,2$)
are fixed in such a way that
the resulting { deviation} from the 
{ underlying pQCD} at high $|{Q^2}| > \Lambda^2$ is:
${\A}_{{\nu}}^{({2 \delta})}(Q^2) - {a(Q^2)}^{{\nu}}  \sim  
( {\Lambda^2}/{{Q^2}} )^{5}$.
The pQCD-onset scale ${M_0}$ is determined so that the
model reproduces the measured (strangeless and massless) $V+A$ tau lepton
semihadronic decay ratio  ${r_{\tau} \approx 0.20}$.

The { underlying pQCD} coupling ${ a}$ is chosen
in {2$\delta$anQCD},
for calculational convenience, in the Lambert-scheme form
{\small
\be
a(Q^2)=-\frac{1}{c_1}\frac{1}{1-c_2/c_1^2+W_{\mp1}(z_{\pm})} \ ,
\label{aptexact}
\ee
}
where: $c_1=\beta_1/\beta_0$; $Q^2=|Q^2|{\rm e}^{i\phi}$, the upper (lower) sign when $\phi \geq 0$ ($\phi < 0$), and
{\small
\be
z_{\pm}=(c_1{\rm e})^{-1} ({|Q^2|}/{\Lambda^2} )^{-\beta_0/c_1}
{\rm exp}\left[i (\pm\pi-{\beta_0} \phi/{c_1} ) \right].
\label{zina2l}
\ee }

\subsection{anQCD models: MPT}
\label{subs:MPT}

\noindent
Nonperturbative physics suggests that the gluon acquires at low
momenta an effective (dynamical)  mass
${m_{\rm gl}} \sim 1$ GeV, and that the coupling
then has the form
{\small
\be
{\A}^{\rm {(MPT)}}(Q^2) = 
a(Q^2+{m_{\rm gl}}^2) \ .
\label{MPT1}
\ee
}
Since ${m_{\rm gl}} > {{\Lambda}_{\rm Lan.}}$,
the new coupling has {no} Landau singularities.


The (generalized) logarithmic derivatives ${\tA}_{ \delta+m}^{\rm {(MPT)}}(Q^2)$ are then uniquely determined
{\small
\be
{\tA}_{ \delta+m}(Q^2) = 
K_{\delta, m} 
\left(\frac{d}{d \ln Q^2}\right)^{m}
\int_0^1 \frac{d \xi}{\xi} {\A}^{\rm {(MPT)}}(Q^2/\xi) \ln^{-\delta}\left(\frac{1}{\xi}\right).
\label{tAnuMPT}
\ee
}

\section{Numerical implementation and results}
\label{sec:num}

Programs of numerical implementation in anQCD models:
\begin{itemize}
\item
for integer power analogs ${\A}_{{n}}(Q^2)$ in {APT} and in ``massive QCD'' \cite{Nest2a,Nest2b}: Nesterenko and Simolo, 2010 (in Maple) \cite{NS1}, and 2011 (in Fortran) \cite{NS2};
\item for general power analogs  ${\A}_{\nu}(Q^2)$ in {FAPT}: Bakulev and Khandramai, 2013 (in Mathematica) \cite{BK};
\item for general power analogs  ${\A}_{\nu}(Q^2)$ in 
{ 2$\delta$anQCD}, {MPT} and {FAPT}: the presented work in Mathematica \cite{arxivCAGC} and Fortran (programs in both languages can be downloaded from the web page: {\bf gcvetic.usm.cl}).
\end{itemize}

The basic relations for the numerical implementation of ${\A}_{{\nu}}$ are:
in {FAPT} Eq.~(\ref{AnuFAPT});
in {2$\delta$anQCD} Eqs.~(\ref{tAnu2d}) and (\ref{AnutAnu});
in {MPT} Eqs.~(\ref{tAnuMPT}) and  (\ref{AnutAnu}).

In Mathematica, ${\rm {Li}}_{-{\nu}}(z)$ is implemented as ${\rm PolyLog}[-\nu,z]$. In Mathematica 9.0.1 it is unstable for $|z| \gg 1$. Therefore, we provide a subroutine Li\_\_nu.m (which is called by the main Mathematica program anQCD.m) and gives a stable version under the name ${\rm polylog}[-\nu,z]$. This problem does not exist in Mathematica 10.0.1. 

In Fortran, program Vegas \cite{vegas} is used for integrations. However, in Fortran, ${\rm {Li}}_{-{\nu}}(z)$ function is not implemented for general (complex) $z$, and is evaluated as an integral Eq.~(\ref{Li_nugen}). 
Therefore, the evaluation of ${\tA}_{{\nu}}$'s in {2$\delta$anQCD} is somewhat more time consuming in Fortran than in Mathematica. Further, more care has to be taken in Fortran to deal correctly with singularities of the integrands.

\begin{figure}[htb] 
\begin{minipage}[b]{.49\linewidth}
\centering\includegraphics[width=80mm]{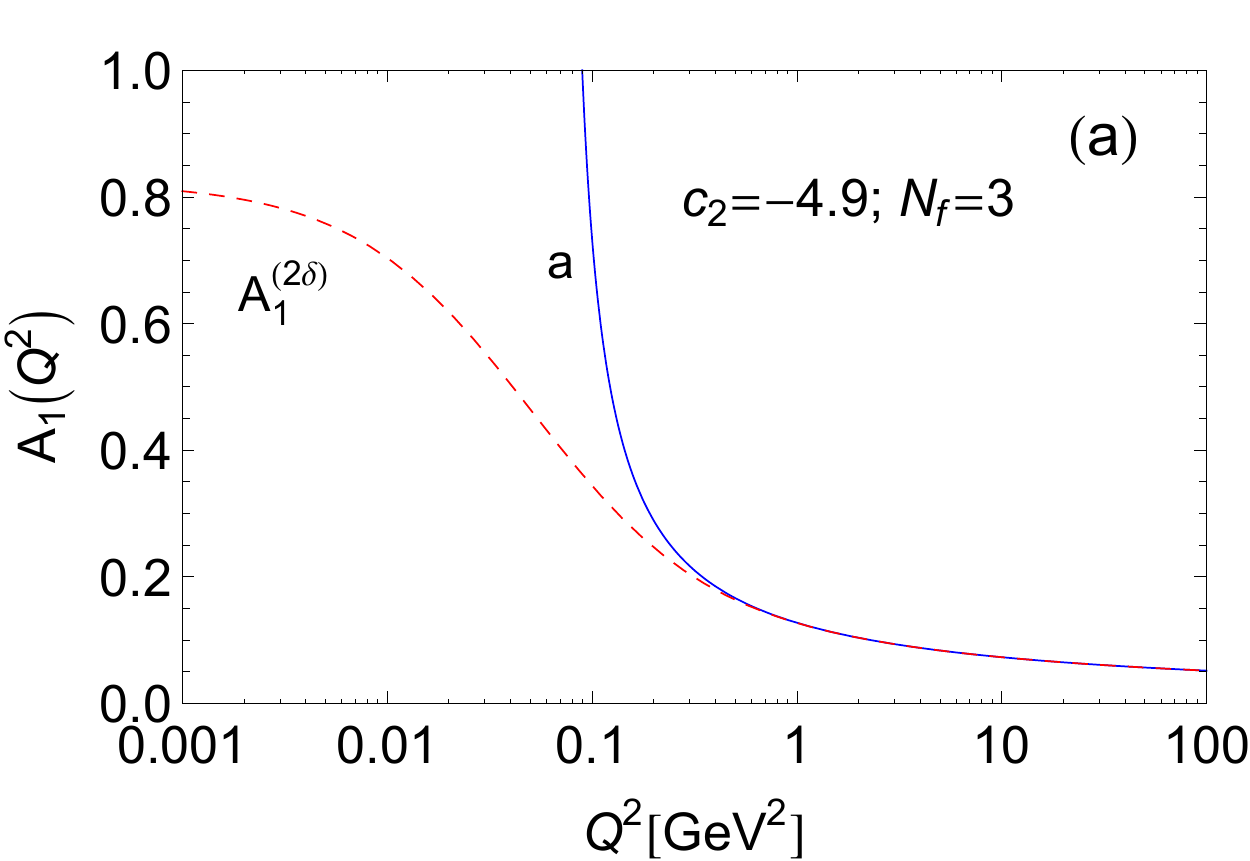}
\end{minipage}
\begin{minipage}[b]{.49\linewidth}
\centering\includegraphics[width=80mm]{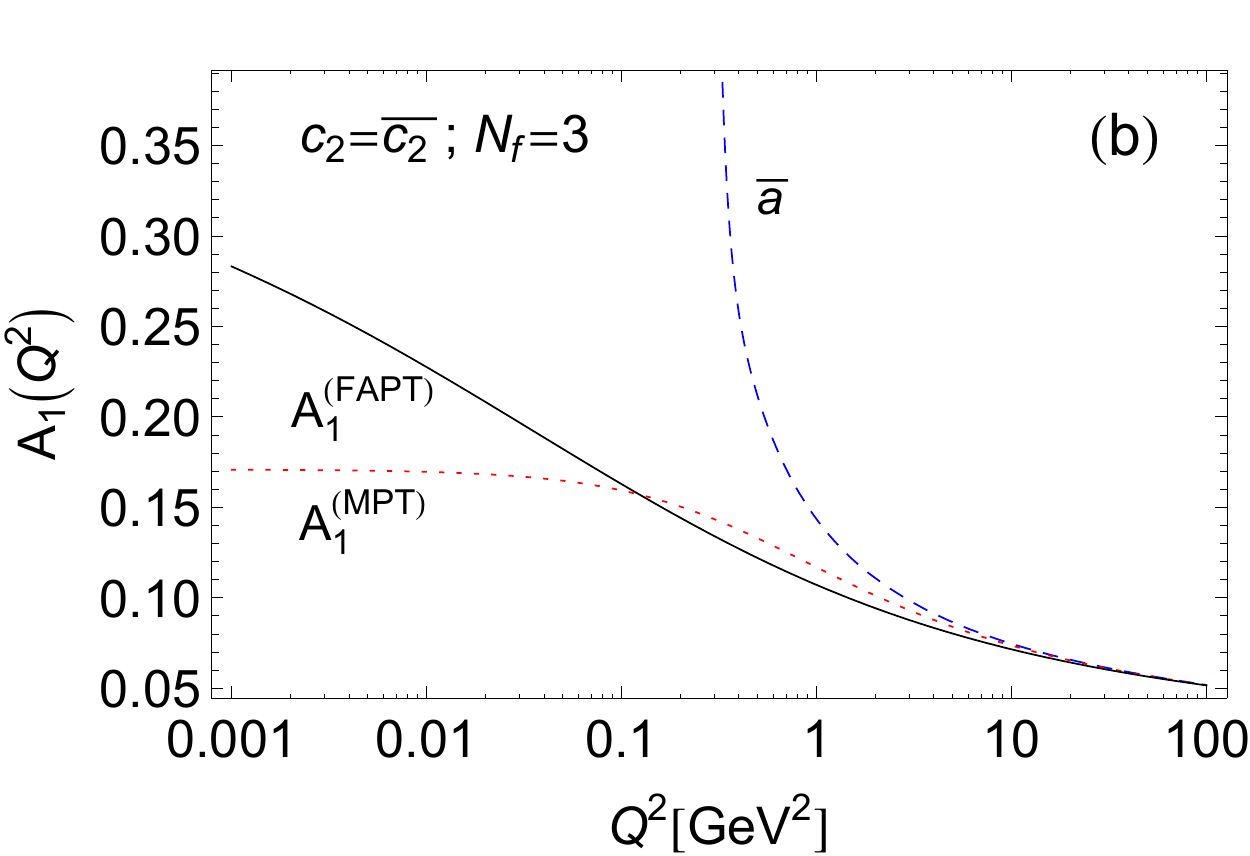}
\end{minipage}
\vspace{-0.4cm}
 \caption{\footnotesize  ${\A}_1 \equiv {\A}$ in 
three anQCD models with ${\nu}=1$ 
 and $N_f=3$, as a function of $Q^2$ for $Q^2 >0$;
the underlying pQCD coupling $a$ is included for comparison: 
 (a) {2$\delta$anQCD} coupling and pQCD coupling, in the Lambert
scheme with $c_2=-4.9$ (and $c_j=c_2^{j-1}/c_1^{j-2}$ for $j \geq 3$);
 (b) {FAPT} and {MPT} in 4-loop $\MSbar$ scheme and with 
$\bL^2_3=0.1 \ {\rm GeV}^2$; {MPT} is with ${m^2_{\rm gl}}=0.7 \ {\rm GeV}^2$.}
\label{figA1L}
 \end{figure}
\begin{figure}[htb] 
\begin{minipage}[b]{.49\linewidth}
\centering\includegraphics[width=80mm]{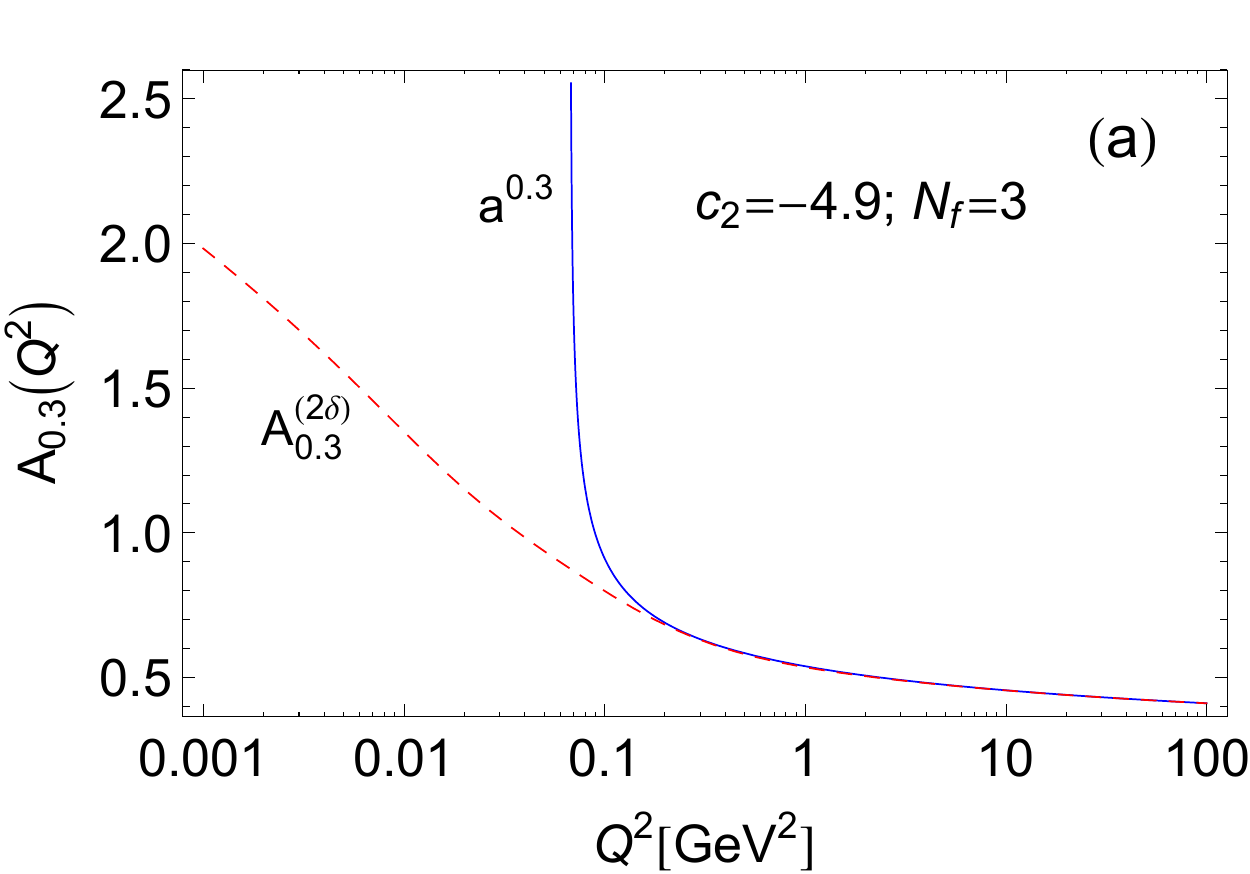}
\end{minipage}
\begin{minipage}[b]{.49\linewidth}
\centering\includegraphics[width=80mm]{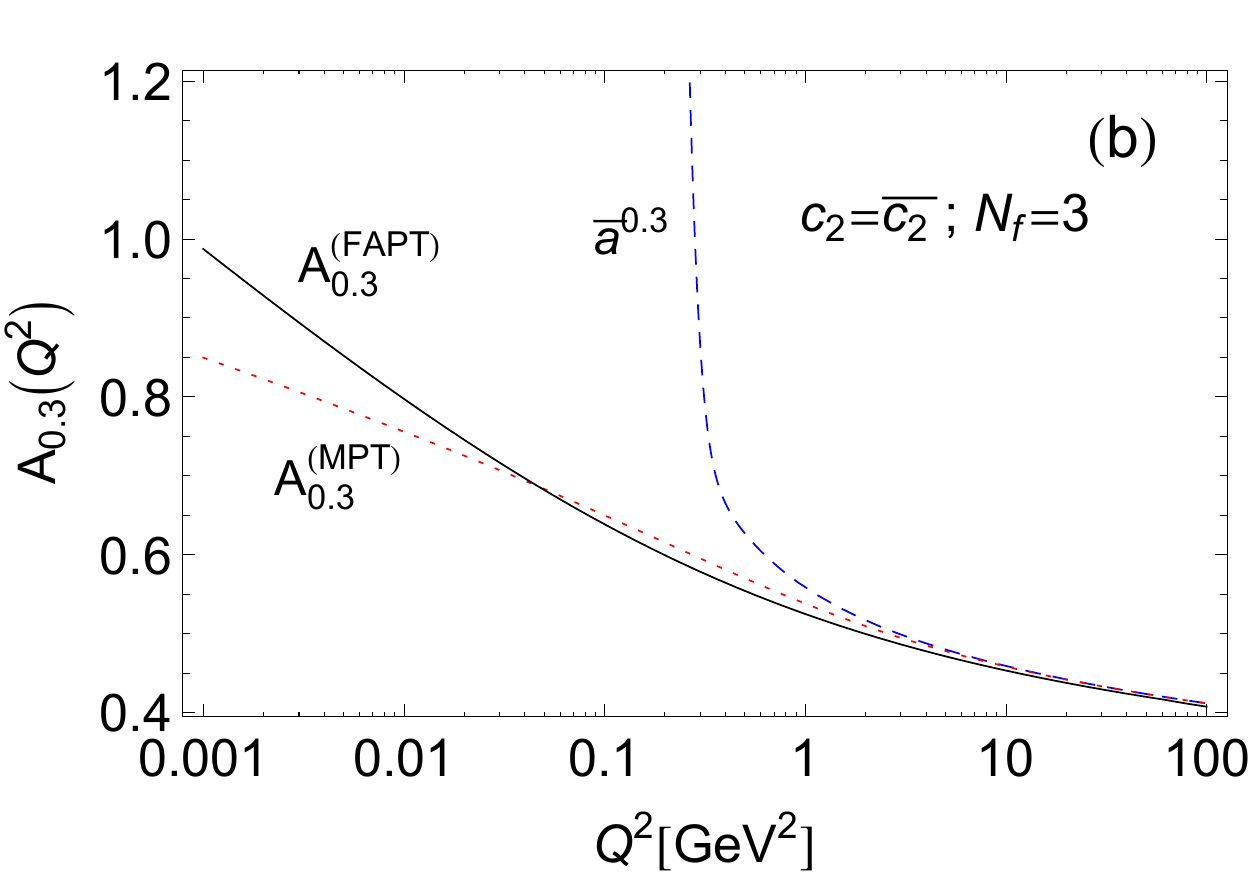}
\end{minipage}
\vspace{-0.4cm}
 \caption{\footnotesize The same as in Fig.~\ref{figA1L}, but 
 with ${\nu}=0.3$ (${\A}_{{\nu}=0.3}$). ${\A}_{0.3}$ is calculated
from ${\tA}_{0.3+m}$ using the relation (\ref{AnutAnu}) 
with $\nu_0=0.3$, and truncation at  ${\tA}_{0.3+4}$ 
in {2$\delta$anQCD},
and at ${\tA}_{0.3+3}$ in {MPT}; and in {FAPT} using Eq.~(\ref{AnuFAPT}).
Figs.~\ref{figA1L} and \ref{figA03L} are taken from \cite{arxivCAGC}.}
\label{figA03L}
 \end{figure}

\section{Main procedures in Mathematica for three analytic QCD models}
\label{sec:proc}

1.) ${\rm AFAPT}N{\rm l}[N_f,{\nu},0,|Q^2|,\Lambda^2,{\phi}]$
gives $N$-loop $(N=1, 2, 3, 4)$
analytic {FAPT} coupling ${\A}_{{\nu}}^{{({\rm FAPT}},N)}(Q^2, N_f)$
with real power index ${\nu}$,
with fixed number of active quark flavors $N_f$, 
in the Euclidean domain [$Q^2 = |Q^2| \exp(i {\phi}) 
\in {\mathcal C}$ and $Q^2 \not< 0$]
{\small
\bea
  {\rm AFAPT}N{\rm l}[Nf,{\nu},0,Q2,L2,{\phi}] &= & 
{{\A}}_{{\nu}}^{({\rm {FAPT}},N)}[Q2=|Q^2|, {\phi}={\rm arg}(Q^2); 
Nf=N_f; L2=\bL^2_{N_f}] 
\nonumber\\
 &&  (N = 1,2,3,4\,;\ Nf = 3,4,5,6). 
\nonumber
\eea
}
2.) ${\rm A2d}N{\rm l}[N_f,M,{\nu},|Q^2|,\phi]$
gives ``$N$-loop'' {2$\delta$anQCD} coupling 
${{\A}}_{{\nu}+M}^{({2\delta})}(Q^2, N_f)$, with
power index ${\nu}+M$ (${\nu} > -1$ and real; $M=0,1,\ldots,N-1$),
with number of active quark flavors $N_f$,
in the Euclidean domain. It is used in the ${\rm N}^{N-1}{\rm LO}$
truncation approach [where in (\ref{AnutAnu}): ${\nu_0} \mapsto {\nu}$ and $n \mapsto M$, and we truncate at ${\tA}_{{\nu}+N-1}$] 
{\small
\bea
  {\rm A2d}N{\rm l}[Nf,M,{\nu},Q2,\phi]
  &=& {{\mathcal A}}_{{\nu}+M}^{{(2\delta)}}[Q2\!=\!|Q^2|, {\phi}\!=\!{\rm arg}(Q^2); Nf\!=\!N_f]\,,
  \nonumber\\
  &&  \!\!\!\!\!\!\!\! \!\!\!\!\!\!\!\!\!\!\!\!\!\!\!\! \!\!\!\!\!\!\!\!\!\!\!\!\!\!
(N= 1,2,3,4,5;\ Nf= 3,4,5,6;\ M=0,1,\ldots,N-1).
\nonumber
\eea
}
3.) ${\rm AMPT}N{\rm l}[N_f,{\nu},Q^2,{m_{\rm gl}}^2,\bL^2_{N_f}]$
gives $N$-loop $(N=1, 2, 3, 4)$ analytic MPT coupling 
${\A}_{{\nu}}^{({\rm {MPT}},N)}(Q^2, {m_{\rm gl}}^2,N_f)$,
with real power index ${\nu}$ ($0<{\nu}<5$)
and with number of active quark flavors $N_f$,
in the Euclidean domain ($Q^2 \in {\mathcal C}$ and $Q^2 \not< 0$)
{\small
\bea
  {\rm AMPT}N{\rm l}[Nf,{\nu},Q2,{M2},L2] & = &
{\A}_{{\nu}}^{(\rm{{MPT}},N)}[Q2\!=\!Q^2 \!\in\! {\mathcal C}; Nf\!=\!N_f; {M2}\!=\!{m_{\rm gl}}^2; L2\!=\!\bL^2_{N_f}] 
  \nonumber\\
&& (N = 1,2,3,4\,;\ Nf = 3,4,5,6)\,;\ 0 < {\nu} < 5). 
\eea
}
\noindent
Examples:

\noindent
Input scale of the {underlying $\MSbar$ pQCD} 
for {FAPT} and {MPT} is $\bL^2_3=0.1 \ {\rm GeV}^2$.
The times are for a typical laptop, using Mathematica 9.0.1;
the first entry in the results is the time of calculation, in $s$.

{\rm In[1]:= $<<$anQCD.m;}


{\rm In[2]:= AFAPT3l[5, 1, 0, $10^2$, 0.1, 0] // Timing}

{\rm Out[2]=} $\{0.382942, 0.0624843\}$

{\rm In[3]:= AMPT3l[5, 1, $10^2$, 0.7, 0.1] // Timing}

{\rm Out[3]=} $\{0.108983, 0.0627726\}$

{\rm In[4]:= A2d3l[5, 0, 1, $10^2$, 0] // Timing}

{\rm Out[4]=}$\{0.768884, 0.0559182\}$

{\rm In[5]:= AFAPT3l[3, 1, 0, 0.5, 0.1, 0] // Timing}

{\rm Out[5]=} $\{0.378943, 0.121853\}$

{\rm In[6]:= AMPT3l[3, 1,  0.5, 0.7, 0.1] // Timing}

{\rm Out[6]=} $\{0.106984, 0.132199\}$

{\rm In[7]:= A2d3l[3, 0, 1,  0.5, 0] // Timing}

{\rm Out[7]=} $\{0.775882, 0.163402\}$


{\rm In[8]:= AFAPT3l[3, 0.3, 0, 0.5, 0.1, 0] // Timing}

{\rm Out[8]=} $\{0.456930, 0.556644\}$

{\rm In[9]:= AMPT3l[3, 0.3, 0.5, 0.7, 0.1] // Timing}

{\rm Out[9]}=$\{0.110983, 0.569473\}$

{\rm In[10]:= A2d3l[3, 0, 0.3, 0.5, 0] // Timing}

{\rm Out[10]}= $\{3.125525, 0.576005\}$

\vspace{0.5cm}

\section{Conclusions}
\label{sec:concl}
We constructed programs, in Mathematica and Fortran,
which evaluate couplings ${\A}_{\nu}(Q^2)$ 
in three models of analytic QCD (FAPT, 2$\delta$anQCD, and MPT). 
These couplings are holomorphic functions (free of Landau singularities)
in the complex $Q^2$ plane with the exception of the
negative semiaxis, and are
analogs of powers $a(Q^2)^{\nu} \equiv (\alpha_s(Q^2)/\pi)^{\nu}$ of the
underlying perturbative QCD. We checked that our results in FAPT model
agree with those of Mathematica program \cite{BK}. 

\vspace{0.5cm}

\noindent
{\bf Acknowledgments}

\noindent
This work was supported by FONDECYT (Chile) Grant No. 1130599 and DGIP 
(UTFSM) internal project USM No. 11.13.12 (C.A and G.C).

\end{document}